\documentclass[%
reprint,
groupedaddress,
bibnotes,
 amsmath,amssymb,
 aps,
 prl,
]{revtex4-1}

\usepackage{graphicx}
\usepackage{epstopdf}
\usepackage{dcolumn}
\usepackage{bm}

\begin{document}

\preprint{APS/123-QED}

\title{Quenching of $\gamma_{0}$ transition results from $p$-wave neutron inducing doorway mechanism}
 \email{gnkim@knu.ac.kr}
 \author{T. F. Wang,$^{1,2}$ X. T. Yang,$^{1}$ T. Katabuchi,$^{6}$ Z. M. Li,$^{1}$ L. H. Zhu,$^{1,2}$\\
 M. W. Lee,$^{3,5}$ G. N. Kim,$^{3\ast}$ T. I. Ro,$^{4}$ Y. R. Kang,$^{4,5}$ M. Igashira,$^{6}$}
\affiliation{%
 $^{1}$School of Physics, Beihang University, Beijing 100191, China\\
 $^{2}$Beijing Advanced Innovation Center for Big Data-based Precision Medicine, Beihang University, Beijing 100191, China\\
 $^{3}$Department of Physics, Kyungpook National University, Daegu 702-701, Korea\\
 $^{4}$Department of Physics, Dong-A University, Busan 604-714, Korea\\
 $^{5}$Research Center, Dong-nam Inst. of Radiological $\&$ Medical Sciences, Busan 619-953, Korea\\
 $^{6}$Research Laboratory for Nuclear Reactors, Tokyo Institute of Technology, Tokyo 152-8550, Japan
 }%
\date{\today}

\begin{abstract}
  $\gamma$-strength function essentially distinguishes the reaction mechanisms of charged particle inelastic and neutron capture reactions, reflecting from the ratios of transition of neutron capture to low-lying states. The extraordinary quenching of $\gamma_{0}$ transition of $p$-wave neutron resonance reaction in 3$s$-region nucleus $^{57}$Fe is observed, for the first time, due to the non-forming 2p-1h doorway state activation which raises the strong intensities of $\gamma_{1}$ and $\gamma_{2}$. The enhancement of low energy $\gamma$-strength function doesn$'$t emerge in the neutron capture reaction resulting from the population of cascade transitions of low-lying states rather than primary transitions. The astrophysical reaction rates that are extracted from total and partial neutron capture cross sections of $^{57}$Fe might be adopted to constrain the abundance of the successive heavier isotopes in $s$-process.

\end{abstract}

\pacs{Valid PACS appear here}
\maketitle


The compound nucleus mechanism, characterized by statistical quantities of level density and $\gamma$-strength function, is dominant in the neutron capture process up to incident neutron energy of several MeV. The resonance regions at low neutron energies, however, show the obvious non-statistical processes, such as potential, valence capture as well as doorway states. The intermediate structure with 2p-1h configuration in doorway states might result in the quenching of the E1 dipole strength [1].

In the neutron resonance region three main non-statistical effects are present, they are the non-constant behavior of the neutron strength function, the anomalous bump at $E_{\gamma}\approx$ 5.5 MeV in the capture $\gamma$-ray spectrum due to the pygmy resonance with an origin from electric transition [2], and the behavior of total radiative widths [3]. The neutron strengths for $s$-wave resonances as a function of the nucleus mass number A have peaks emerging at A=55 and A=150, however, non-enhancement arise for $p$-wave neutron resonance in these mass region. These peaked $s$-wave neutron strength mass regions corresponding to nuclei where the 3$s$ and 4$s$ states are slightly unbound, are normally called 3$s$ and 4$s$ region. The pygmy resonance originates from electric transition. It was declared that the spectra of high energy $\gamma$-rays in nuclei in the 3$s$ region, manifested a higher radiation as should be accounted for by an $E_{\gamma}^{3}$ energy dependence for the transition probability and an exponential level-density energy dependence, which are assumed in the statistical model. As the partial reduced widths follow the Porter-Thomas distribution, the total radiative widths are expected to have a very narrow distribution. However, Cameron [3] fitting the total radiation widths for all nuclei, taking into account level-density effects, found a small residual variation with A. A more pronounced peak was noticed in the mass region between A=145 and 275, which coincides with 4$s$ region. In addition, fluctuations in the total radiative cross-sections for different resonances were frequently found to be quite large, indicating that the number of degrees of freedom was less than expected by the statistical model. Therefore, the mass regions where non-statistical effects are more pronounced are 3$s$ and 4$s$ regions.

A comprehensive theoretical framework of the neutron capture process was developed, where the statistical model was associated to single or semi-collective interactions to account for non-statistical experimental evidences. Lane and Lynn [4] first identified three main components of contributions to the capture process: the compound nucleus (or resonance internal), channel(or resonance external) and direct capture (hard sphere potential and distant resonance). Subsequent developments included the possibility of a semi-direct capture in the giant dipole resonance in order to interpret the observed discrepancies at high incident neutron energies (above 5 MeV) [5]. All the neutron capture processes are presented and classified into a non-resonant and a resonant contribution [6]. In the resonance contribution, the channel capture has been split into its valence and doorway parts. The valence capture achieves preferentially single particle final states where the valence neutron can maintain a radiative transition without perturbing the core. In the doorway capture only, a two-particle one-hole (2p-1h) configuration is formed, the radiative decay can be generated either by the neutron transition in the excited nucleus or by a particle-hole annihilation. When the incident particle undergoes many interactions and a large number of particle-hole configurations are produced, the statistical model must be utilized to take into account the intrinsic pattern.

As regards the $\gamma$-strength function, a great of efforts have been made to insight into the intrinsic sense from various formulations. A model-independent approach reflecting $\gamma$-strength function was imperatively carried out [7]. The ratio of photon strength function with two different primary transitions from the same initial excitation energy $E_{i}$ to discrete low-lying levels of the spin and parity at energies $E_{L1}$ and $E_{L2}$, is formulated as,

\begin{equation}
\begin{split}
R=\frac{f(E_{i}-E_{L1})}{f(E_{i}-{E_{L2}})}=\frac{N_{L1}(E_{i})(E_{i}-E_{L2})^{3}}{N_{L2}(E_{i})(E_{i}-E_{L1})^{3}},
\end{split}
\end{equation}

where $N_{L1}(E_{i})$ and $N_{L2}(E_{i})$ is the intensities of primary transitions from the neutron capture state to the low-lying states, their transition energies are equal to $E_{i}-E_{L1}$ and $E_{i}-E_{L2}$. In this letter, it is noticed that the ratio $R$ have a pronounced enhancement at the only $p$-wave neutron resonance with respect to the almost uniformed distribution of the mixed neutron resonances (narrow $p$-wave resonances superpose on a wide $s$-wave resonance) in the 3$s$-region nucleus of $^{57}$Fe. This enhancement probably originates from the doorway mechanism formed by  capture-neutron coupling to the core-neutron p-h configuration, dominated by strong E1 transitions of $\gamma_{1}$ and $\gamma_{2}$. However, the $\gamma_{0}$ transition seems tremendously suppressed due to none of the formation of doorway mechanism for the induction of $p$-wave neutron resonance.

The measurement was carried out at the 3 MV Pelletron in the Tokyo Institute of Technology [8]. Proton beam with energy of 1.903 MeV hitting on $^{7}$Li target, which was made by evaporating metallic lithium on a copper disk, produced 2-90 keV continue neutrons. The shape of this continuum neutron spectrum is much close to the distribution of Maxwellian-Boltzmann [8]. Hence this neutron generator can give a realistic condition of the astrophysical environment of slow neutron-capture ($s$-) process. The nuclear reaction rate close to the reality could be constrained from the experimental total and partial neutron capture cross sections. The energy of incident neutrons on a $^{57}$Fe enriched sample with a thickness of 0.373 mm was measured by means of the time-of-flight (TOF) method with a mini $^{6}$Li-glass scintillation detector located 30 cm from the neutron target. A large anti-Compton spectrometer [8] consisting of a central main NaI(Tl) detector with a 15.2 cm diameter by 30.5 cm and a circular arrangement of NaI(Tl) detectors with a 33.0 cm outer diameter by 35.6 cm was used for detecting $\gamma$-rays from neutron capture reaction. A low peak-efficiency less than 3.5$\%$ can essure to access only one $\gamma$-ray in each cascade decay. A good signal-to-noise ratio was obtained due to the powerful shield composed of borated paraffin, borated polyethylene, cadmium, and potassium free lead, besides, a $^{6}$LiH cylinder which absorbed effectively the neutrons scattered by the sample was added in the collimator of the spectrometer shielding shell. The spectrometer was placed at an angle of 125$^{\circ}$ with respect to the proton beam direction by kinematics, because the second Legendre polynomial is zero at this angle, the differential measurement at this angle gave approximately the integrated $\gamma$-ray spectrum for the dipole transition. The distance between the center of neutron capture sample and the front surface center of main NaI(Tl) detector is fixed to be 78.5 cm.

The neutron capture cross-sections of certain range ($i$) of neutron energy are extracted from the capture yield ($Y_{i}$), the sample thickness ($n$), the number of incident neutron on sample ($\phi_{Fe}$) that are deduced from the relative measurement of $\phi_{Au}$ for the standard Au sample normalized by neutron monitor ($M$). $Y_{i}$ is equal to the ratio of weighted $\gamma$-spectrum area ($\sum_{I}W(I)S_{i}(I)$) to the neutron binding energy including the averaged neuron energy ($B_{n}+<E_{n}>$) [9].

\begin{equation}
\begin{split}
<\sigma>_{i}^{Fe}&=\frac{Y_{i,Fe}}{C_{i,Fe}n_{Fe}\phi_{Fe}}\\
                 &=\frac{C_{i,Au}}{C_{i,Fe}}\cdot\frac{n_{Au}}{n_{Fe}}\cdot\frac{M_{Au}}{M_{Fe}}\cdot\frac{B_{n,Au}+<E_{n,Au}>}{B_{n,Fe}+<E_{n,Fe}>}\\
                 &\cdot\frac{\sum_{I}W(I)S_{i,Fe}(I)}{\sum_{I}W(I)S_{i,Au}(I)}\cdot<\sigma>_{i,Au},
\end{split}
\end{equation}

Where $C_{i}$ are correction factors including the effects of neutron multiple-scattering and self-shielding, $\gamma$-ray absorption and scattering in sample, the angle distribution of neutron flux.

Since the number and individual energies of emitted $\gamma$-rays are different for the cascade mode by mode, the efficiency for detecting capture events depends on decay modes. The pulse height weighting method was used to obtain the detection efficiency that is proportional to the energy dissipated in the detector. The net $\gamma$-ray spectra were extracted by subtracting the background spectrum normalized with the digital gate widths of the foreground spectrum. The unfolded net $\gamma$-ray spectrum via the response function matrix of the spectrometer is shown in Fig. 1, where the strong primary transitions from the neutron capture states to the ground state ($E_{x}$ = 0 MeV, $J^{\pi}$ = 0$^{+}_{GS}$), the first excited state (0.811 MeV, 2$^{+}_{1}$) and the second excited state (1.675 MeV, 2$^{+}_{2}$) are observed around 10.09, 9.28, and 8.42 MeV, respectively. The obvious peaks around 0.8 and 1.7 MeV are produced mainly due to the cascade transitions from the first (2$^{+}_{1}$) and the second excited state (2$^{+}_{2}$) to the ground state (0$^{+}_{GS}$).

\begin{figure}
\includegraphics[width=9.6cm]{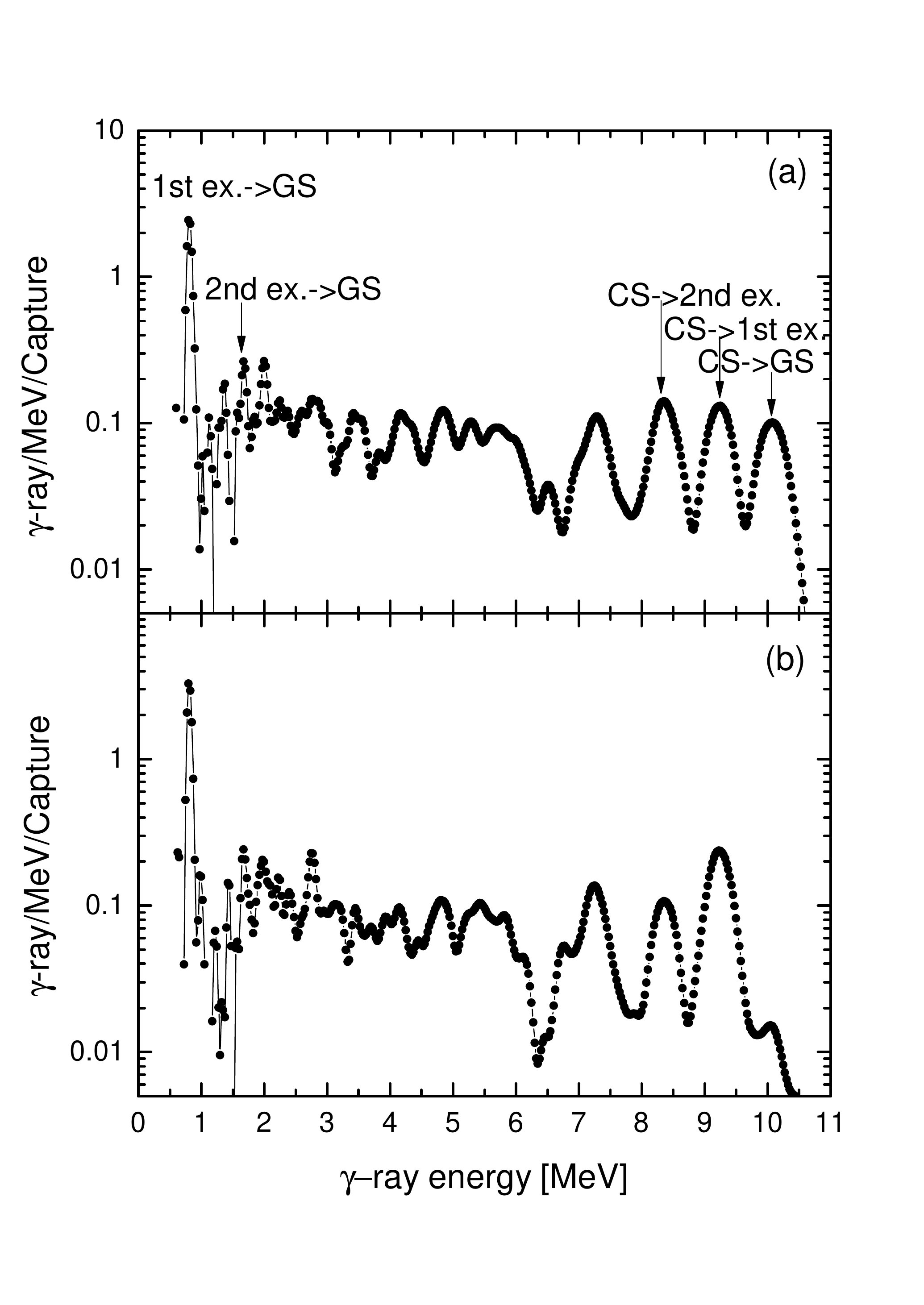}
\caption{\label{fig:pdfart}The unfolded net $\gamma$-ray spectra of $^{57}$Fe(n,$\gamma$)$^{58}$Fe for: (a)${\textless}E_{n}{\textgreater}$ = 21.56 keV, (b)${\textless}E_{n}{\textgreater}$ = 17.59 keV.}
\end{figure}

The intensity of primary transition is proportion to the $\gamma$-strength function of $f(E_{i}-E_{Lj})$, the transition energy of $(E_{i}-E_{Lj})^{3}$ as well as $\sum_{J\pi}\sigma_{J\pi}(E_{i})$ that is the cross section for populating the levels with the given spin and parity at excitation energy $E_{i}$ [7]. The mostly populated neutron capture state of 1$^{-}$ for $^{58}$Fe is created when incident $s$-wave neutron with single Fermi particle state couples to the low-lying states of $^{57}$Fe. Since $\sigma_{1^{-}}(E_{i})$ keeps as a constant for different modes of $\gamma$-ray decay cascade with the dominate E1 transitions. Therefore, the ratio of $\gamma$-strength function $R$ in Equ. (1) can be extracted from the intensities and energies of primary transitions from neutron capture state to the low-lying states with certain spin and parity. It is essentially significant to investigate the intrinsic characteristics of $\gamma$-transitions for $s$- and $p$-wave neutron, especially for the primary transitions from a common initial state to final states with the same spin and parity. Fig. 2 shows the distributions of $R_{2_{1}^{+}/0_{GS}^{+}}$, $R_{2_{2}^{+}/0_{GS}^{+}}$ and $R_{2_{1}^{+}/2_{2}^{+}}$ depending on neutron resonance energy reflecting $s$- and $p$- wave neutron¡¯s remarkable coupling feature. A distinguish enhancement in $R_{2_{1}^{+}/0_{GS}^{+}}$ for $p$-wave resonance is exhibited comparatively with the slightly varied amplitude of the $s$ and $p$-wave superimposing resonances. $R_{2_{2}^{+}/0_{GS}^{+}}$ displays a similar behavior as that of $R_{2_{1}^{+}/0_{GS}^{+}}$, but with a slightly weaker strength. The higher ratio of $p$-wave with respect to $s$-wave in $R_{2_{1}^{+}/2_{2}^{+}}$ distribution indicates the $\gamma$-strength function from the initial capture state induced by $p$-wave neutron are much changeable than that of $s$-wave neutron to the low-lying states with the same spin and parity.

\begin{figure}
\includegraphics[width=9.6cm]{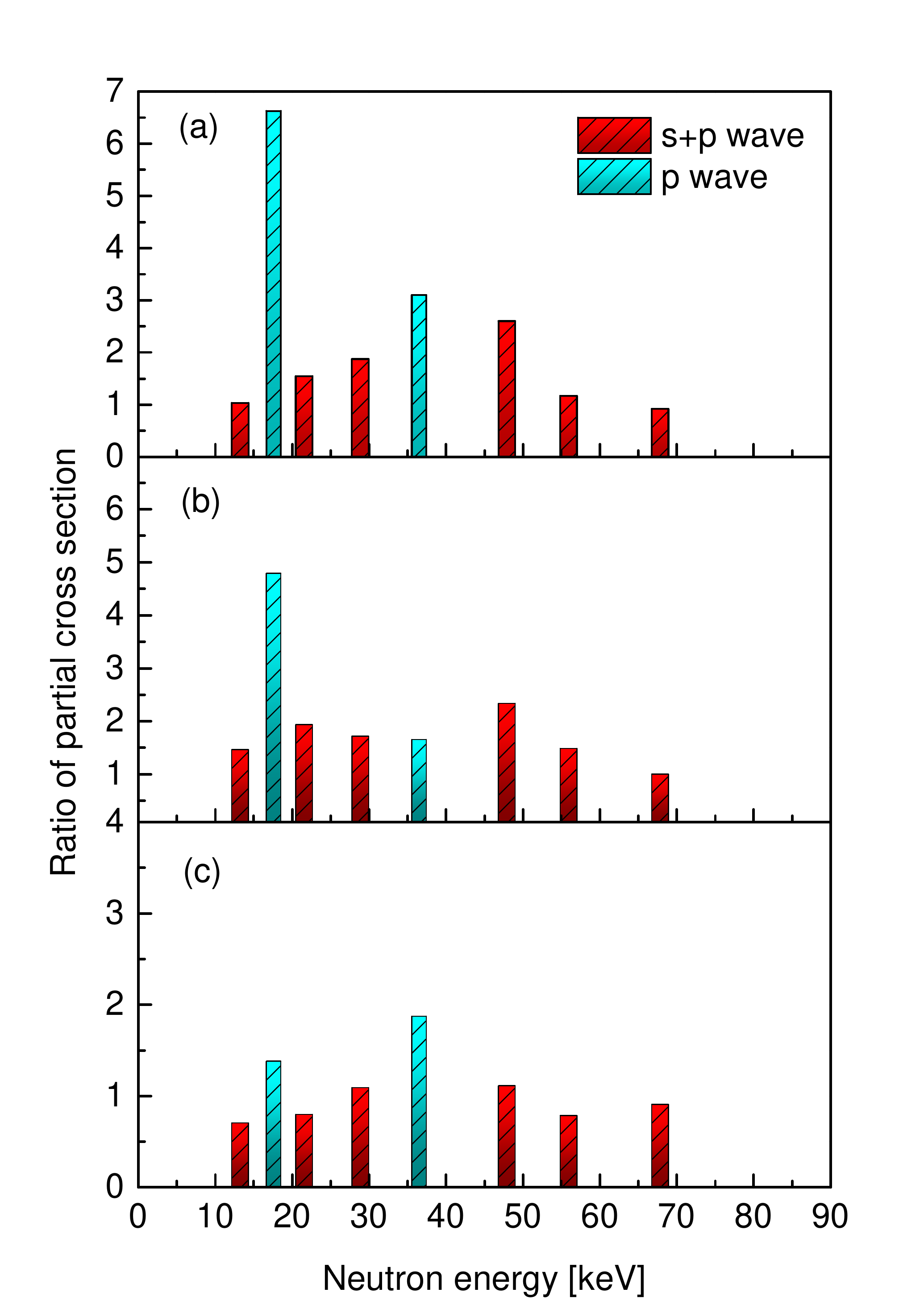}
\caption{\label{fig:pdfart}The distributions of ratio of the primary transition intensities : (a) $R_{2_{1}^{+}/0_{GS}^{+}}$, (b) $R_{2_{2}^{+}/0_{GS}^{+}}$ and (c) $R_{2_{1}^{+}/2_{2}^{+}}$ depending on neutron resonance energies.}
\end{figure}

Neutron capture resonance initial and final states can be represented as single neutron states coupled to the target ground state as well as quadrupole and octupole excitations. These collective states are coherent superpositions of 2p-1h qusiparticle states [10, 11]. If certain p-h configurations are dominant, they may occur with unperturbed energies much greater than those of the collective states [10, 11]. In 3$s$-region possible 2p-1h configurations which can decay by E1 annihilation are (2p$_{1/2}$; 2s$_{1/2}^{-1}$), (2p$_{1/2}$; 1d$_{3/2}^{-1}$), (1f$_{5/2}$; 1d$_{3/2}^{-1}$) neutron particle-hole pairs coupled to the 2p$_{1/2}$, 2p$_{3/2}$ or 1f$_{5/2}$ neutron orbit. The energies of p-h components are 8-10 MeV according to the neutron binding energies of each level, which are exactly consistent with the observed $\gamma$-ray energies of primary transitions ($E_{\gamma_{0}}$ = 10.09 MeV),($E_{\gamma_{1}}$ = 9.28 MeV) and ($E_{\gamma_{3}}$ = 8.42 MeV) in $^{57}$Fe(n, $\gamma$)$^{58}$Fe reaction. If the annihilation energy of paired neutron-hole for the 2p-1h configuration is comparable to the observed $\gamma$-ray energies, the corresponding doorway states are expected to be strongest.

\begin{figure}
\includegraphics[width=8.7cm]{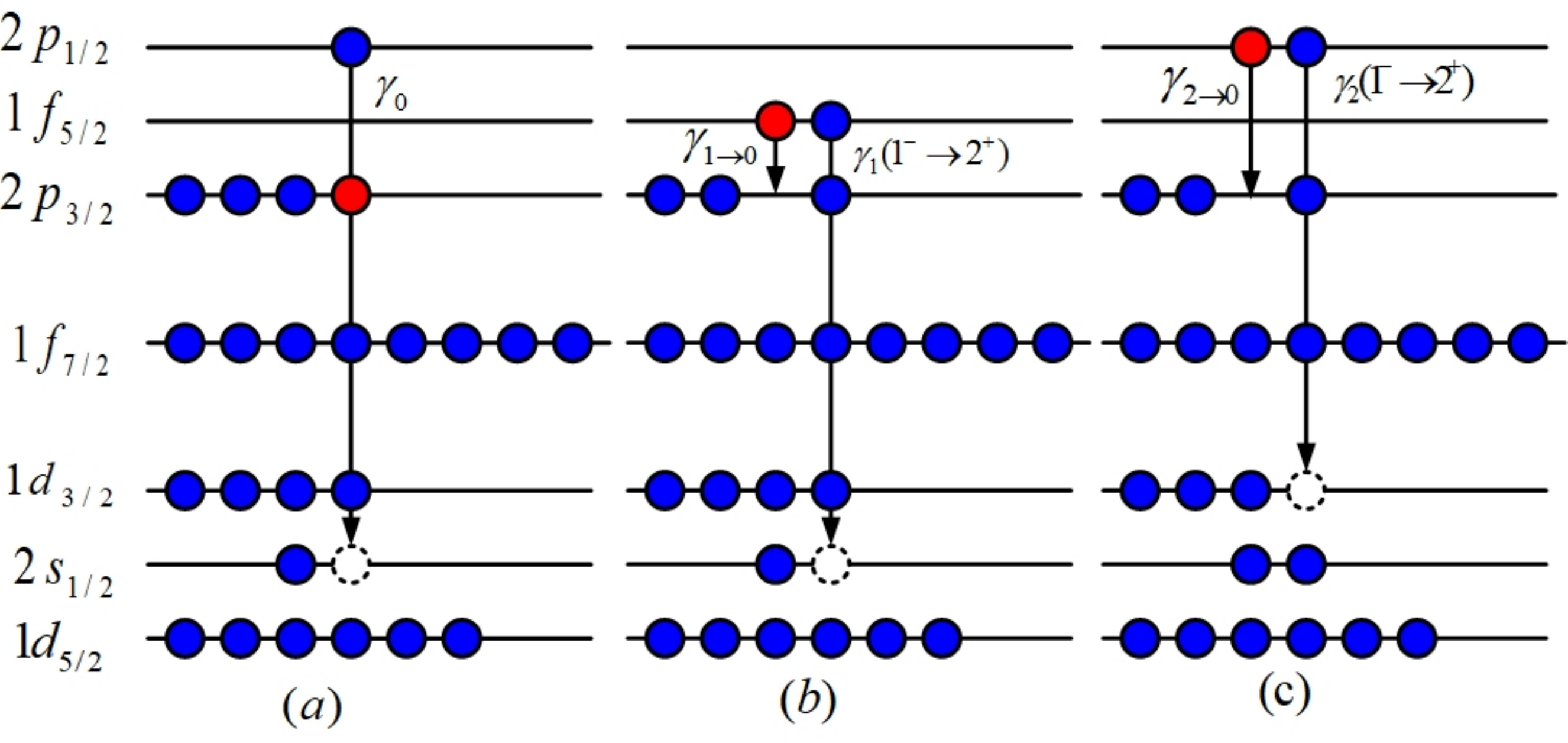}
\caption{\label{fig:pdfart}The configurations of doorway states correspond to the primary transitions of: (a) $1_{CS}^{-}\rightarrow0^{+}$, (b) $1_{CS}^{-}\rightarrow2_{1}^{+}$, (c) $1_{CS}^{-}\rightarrow2_{2}^{+}$ in $^{58}$Fe.}
\end{figure}

The configurations of doorway states correspond to the primary transitions of $1_{CS}^{-}\rightarrow2_{1}^{+}$, $1_{CS}^{-}\rightarrow2_{2}^{+}$ in $^{58}$Fe are neutron (2p$_{3/2}^{3}$, 1f$_{5/2}^{2}$, 2s$_{1/2}^{-1}$) and (2p$_{3/2}^{3}$, 2p$_{1/2}^{2}$, 1d$_{3/2}^{-1}$), respectively, as shown in Fig. 3, where $1_{CS}^{-}$ is the $s$-wave neutron initially capture state. Several narrow $p$-wave resonances superpose on the wide $s$-wave resonance, therefore $s$+$p$ wave is used to distinguish the independent $p$-wave resonance in Fig. 2. The remarkable enhancement of $R_{2_{1}^{+}/0_{GS}^{+}}$ and $R_{2_{2}^{+}/0_{GS}^{+}}$ for $p$-wave neutron resonance essentially result from the quenching of the intensity of $\gamma_{0}(1_{CS}^{+}\rightarrow0_{GS}^{+})$ which is caused by the non-doorway excitation mechanism due to the capture-neutron filling in 2p$_{3/2}$ to produce the full sub-shell and non-coupling with the neutron p-h pair of (2p$_{1/2}$, 2s$_{1/2}^{-1}$). On the other hand, the transition intensity of $\gamma_{0}(1_{CS}^{+}\rightarrow0_{GS}^{+})$ for $p$-wave neutron is weaker than that of $\gamma_{0}(1_{CS}^{-}\rightarrow0_{GS}^{+})$ for $s$-wave neutron due to the low ratio of magnetic to electric dipole transition probability of $\lambda$(M1):$\lambda$(E1). Since M1 transitions will be hindered with respect to E1 transitions, the $E_{\gamma}^{(2l+1)}$ energy dependence will result in the significant difference between the $s$- and $p$- wave total radiative widths, also the difference in strength. The averaged width of $\textless\Gamma_{\gamma}(s)\textgreater\sim3\textless\Gamma_{\gamma}(p)\textgreater$ was found in the 3$s$ region [10, 11], where there are correlations between the reduced neutron widths and radiative widths of $s$-wave resonance. The doorway excitation mechanism with neutron 2p-1h configurations in $\gamma_{1}(1_{CS}^{-}\rightarrow2_{1}^{+})$ and $\gamma_{2}(1_{CS}^{-}\rightarrow2_{2}^{+})$ cause the strong transitions and high intensities to populate the final states of $2_{1}^{+}$ and $2_{2}^{+}$. The transitions of $2_{1}^{+}\rightarrow0_{GS}^{+}$ and $2_{2}^{+}\rightarrow0_{GS}^{+}$ in cascade decays are interpreted by the capture-neutron jumping from 1f$_{5/2}$ and 2p$_{1/2}$ to 2p$_{3/2}$ forming full 2p$_{3/2}$ sub-shell.

\begin{figure}
\includegraphics[width=9.6cm]{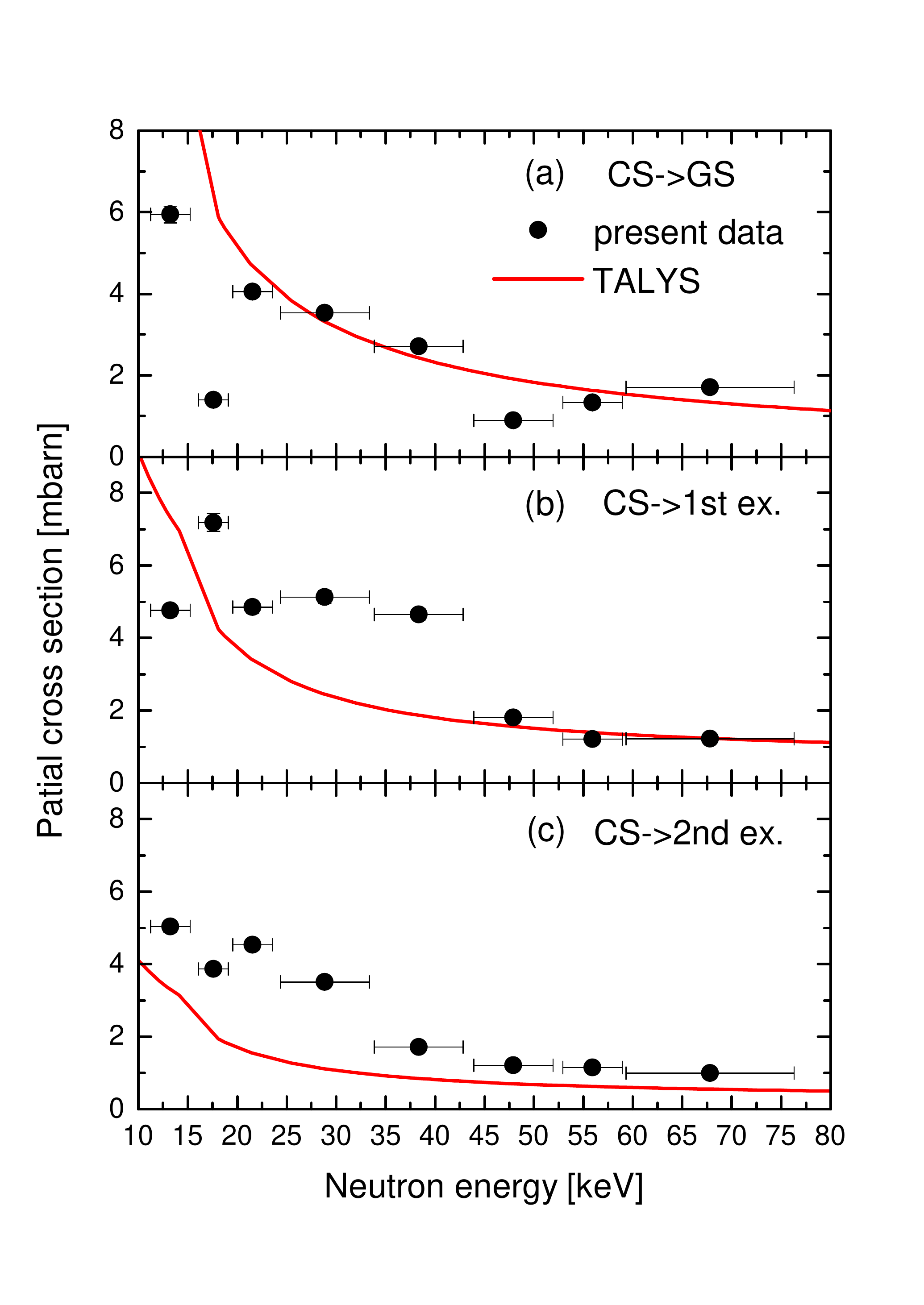}
\caption{\label{fig:pdfart} The experimental partial cross sections to the three lowest $0^{+}$(CS$\rightarrow$GS) and $2^{+}$ (CS$\rightarrow$1st ex. and 2nd ex.) are compared to the calculation using the statistical model code TALYS [16].}
\end{figure}

\begin{figure}
\includegraphics[width=9.6cm]{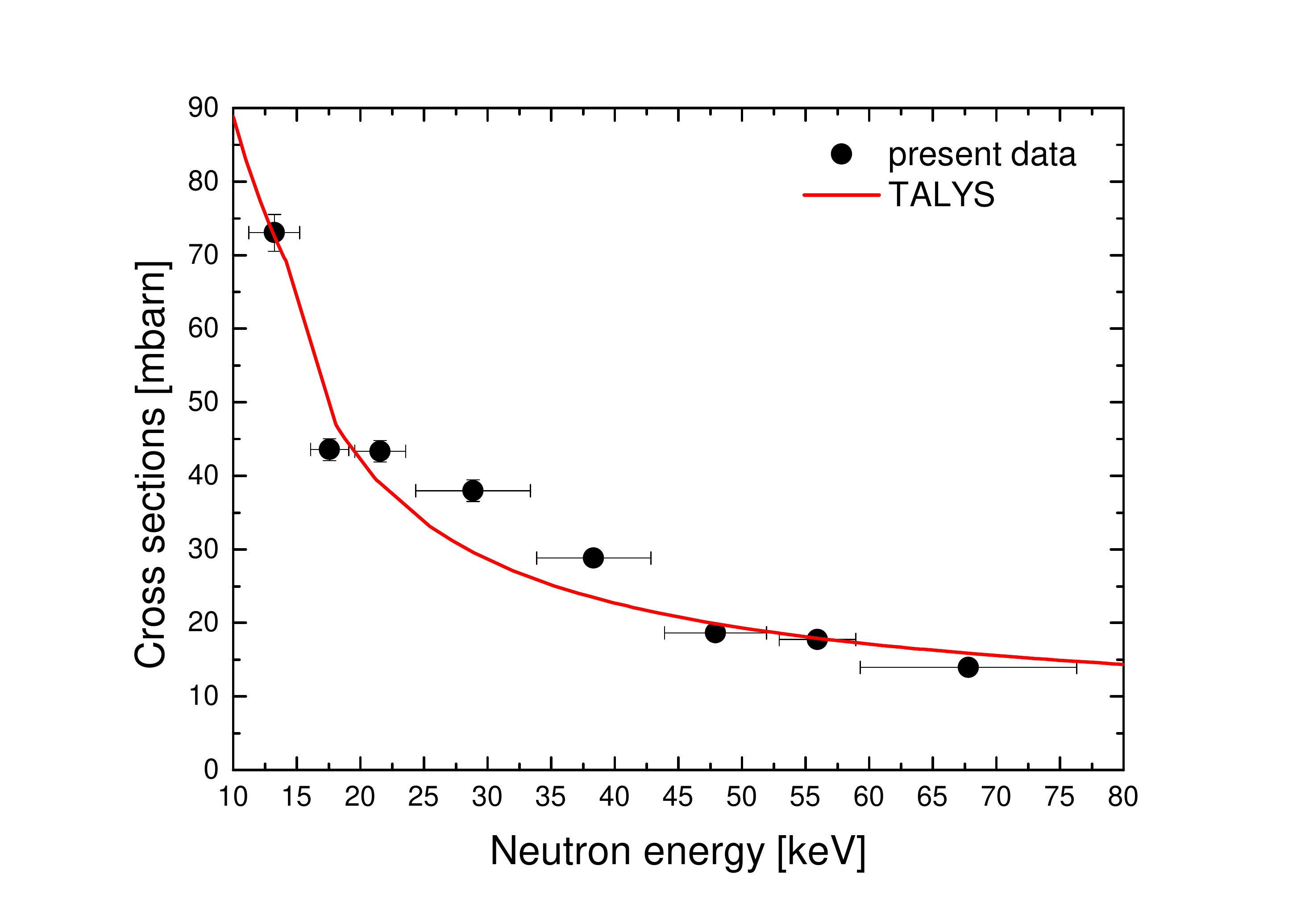}
\caption{\label{fig:pdfart} The experimental total cross sections are compared to the calculation using the statistical model code TALYS [16]}
\end{figure}

Studies of $\gamma$-strength function mostly rely on a great deal of data accumulated in neutron capture reaction [7], from photon scattering facilities. In addition, charge particles induced ($^{3}$He, $^{3}$He$\gamma$) and ($^{3}$He, $\alpha\gamma$) reactions were utilized to determined $\gamma$-strength function at low $\gamma$-ray energies [12, 13], where the pronounced enhancement of $\gamma$-strength function below 3 MeV was observed due to the increase of B(M1) strength of low energy M1 transitions. This consequence was confirmed by (d, p) reaction on $^{95}$Mo target [7]. The enhancement strangely, however, didn$'$t present in neutron capture experiments [14, 15]. This inconsistency highly motivates the studies of the reaction mechanism from $\gamma$-strength function of neutron capture at neutron energy of keV region where the non-statistical and statistical processes are successively covered. From the doorway mechanism point of view, neutron capture reaction in keV region is dominated by 2p-1h excitation, where the $\gamma$-rays are emitted mostly by the annihilation of neutron particle-hole pair with $E_{\gamma}\sim$ 8-10 MeV. The core excitation results in the continuum region from the intermediate $\gamma$-ray energies in the spectrum of Fig. 1. However, the low-energy $\gamma$-rays mostly come from the cascade decay of low-lying states, rather than from the primary transition to the high-lying state. It is distinct that in the inelastic charge particle induced reaction ($^{3}$He, $^{3}$He$\gamma$) where the primary M1 transitions with $E_{\gamma}\textless$3 MeV can be populated to a great extent.

The experimental total and partial cross sections to the three lowest $0^{+}$(CS$\rightarrow$GS) and $2^{+}$ (CS$\rightarrow$1st ex. and 2nd ex.) are compared to the calculation using the statistical model code TALYS [16], as shown in Fig. 4 and Fig. 5. The microscopic level densities of temperature dependent Hartree-Fock-Bogoliubov with Gogny force, and E1 strength function from the Hartree-Fock-BCS with a scaling factor equal to 0.85, and the active JLM neutron potential are adopted in the calculations. The E1 strength function and level density of $^{57}$Fe has recently been extracted from previous measurements [13, 17, 18, 19]. According to the Brink-Axel hypothesis, this $\gamma$-strength function should result in a consistent description compared to the experimental partial cross sections, when it is utilized as an input for TALYS. However, the non-ignorable discrepancies emerged between calculations and experimental data when it combined to other nuclear ingredients in the model. The partial cross sections calculated using level densities of temperature dependent Hartree-Fock-Bogoliubov and strength function from the Hartree-Fock-BCS show in general a similar behavior. The calculation for total cross section exhibits a perfect consistency with experimental data.

\begin{figure}
\includegraphics[width=9.6cm]{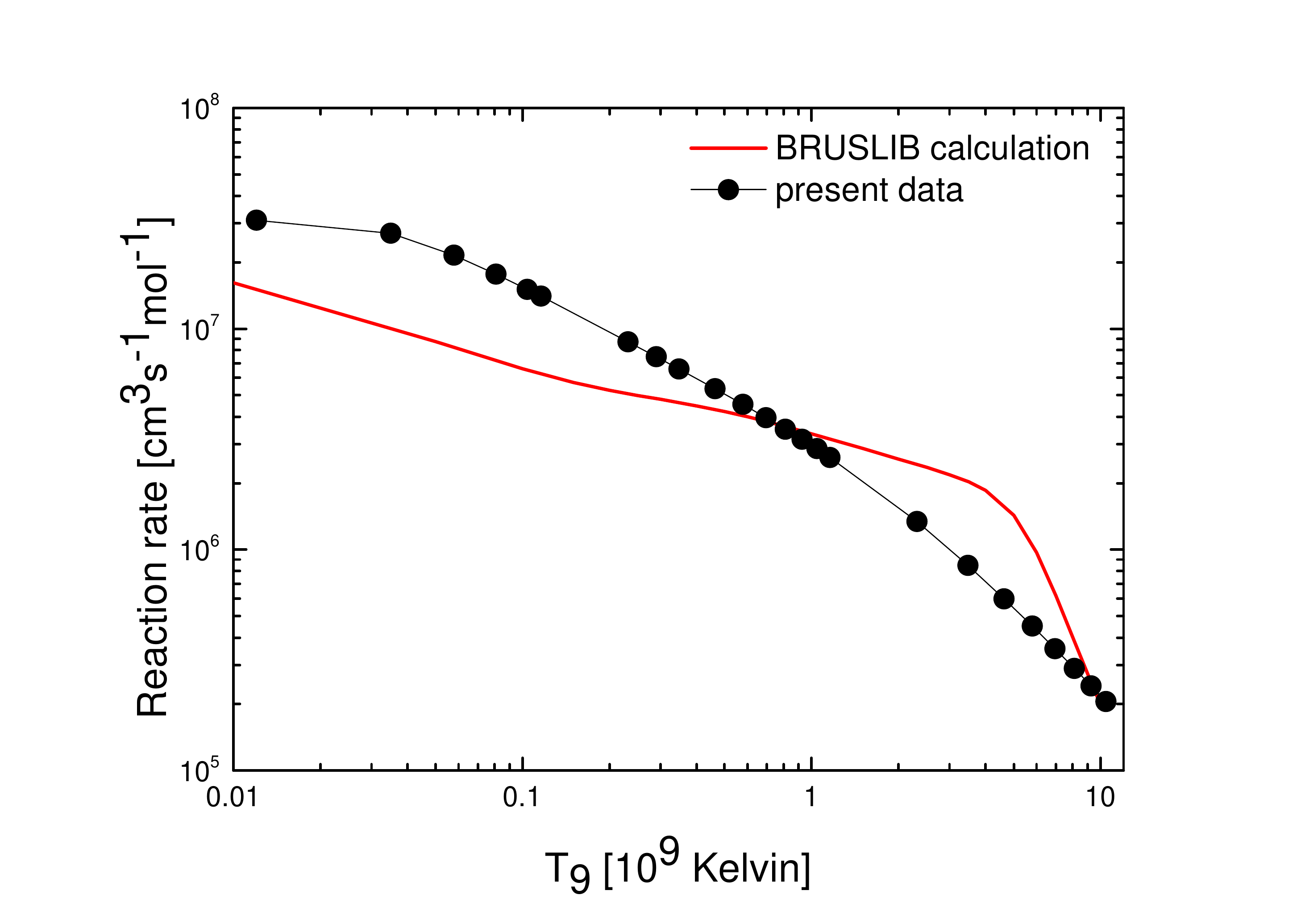}
\caption{\label{fig:pdfart} The astrophysical reaction rates are compared to the calculation in BRUSLIB [23]}
\end{figure}

Most of the heavier elements beyond iron and nickel were created in the slow neutron capture process ($s$-process) in stars. $s$-process occurs at moderate neutron densities around 10$^{8}$ cm$^{-3}$, where neutron captures and subsequent radioactive $\beta$-decays come into being the isotopes along the line of stability. The weak component of $s$-process is responsible for forming the abundances between Fe and Sr, and takes place in massive stars (M $\textgreater$ 8-10 M$_{\bigodot}$) which later explode as core collapse supernovae [20]. The weak $s$-process is first activated as the end of He core burning at temperatures around 0.3 GK (10$^{9}$ Kelvin), and later during C shell burning at temperatures around 1 GK, where neutrons are produced by $^{22}$Ne($\alpha$, n) reactions. The cross section of a single isotope can affect the abundances for a number of successively heavier isotopes. It was found that this effect is especially distinct in the mass region of the weak component of the $s$-process [20].

The Maxwellian-averaged neutron capture cross section for temperature $T$ is given by [21]

\begin{equation}
\begin{split}
\frac{<{\sigma}v>_{kT}}{v_{T}}=\frac{2}{\sqrt{\pi}}\frac{1}{(kT)^{2}}\int_{0}^{\infty}E_{c.m.}{\sigma}exp(\frac{-E_{c.m.}}{kT})dE_{c.m.},
\end{split}
\end{equation}

where $k$ is Boltzmann$'$s constant, $v_{T}=(2kT/\mu)^{1/2}$ and $\mu$ is the reduced mass in the entrance channel, $E_{c.m.}$ is the center-of-mass energy of neutron. The reaction rate $N_{A}\textless{\sigma}v\textgreater$ is obtained with the unit of cm$^{3}$ sec$^{-1}$ mole$^{-1}$ in Fig. 6, where $N_{A}$ is Avogadro$'$s number. The thermal neutron and higher energy neutron capture cross sections in the previous measurements [22] are adopted together with the present data for the above folding integration. The reaction rate are compared to the calculation in BRUSLIB [23], Larger deviation of reaction rates with respect to the calculation is found by factors from two to three in the low temperature region with T$_{9}\textless$ 0.8 GK, however, a good agreement is obtained nearby T$_{9}$ = 1 GK. The reflection point showing in the calculation doesn$'$t emerge in the high temperature region, but decrease with a smooth tendency. The astrophysical reaction rates that are extracted from total and partial neutron capture cross sections of $^{57}$Fe might be adopted to constrain the abundance of the successive heavier isotopes in $s$-process.

In summary, $\gamma$-strength functions of $\gamma_{0}$, $\gamma_{1}$ and $\gamma_{2}$ of the $^{57}$Fe(n, $\gamma$)$^{58}$Fe reaction were studied by the ratio of transition intensities via in-beam $\gamma$-ray spectroscopy. The quenching of transition intensity of $\gamma_{0}$ of $p$-wave neutron resonance is observed with respect to the high intensities of $\gamma_{1}$ and $\gamma_{2}$ due to the effect of doorway mechanism. The enhancement of low energy $\gamma$-strength function does$'$n emerge in the neutron capture reaction resulting from the population of cascade transitions of low-lying states rather than primary transitions. Total and partial cross section are extracted in the measurement, and compared with statistical model TALYS of Hauser-Feshbach calculation, a general consistence is obtained. The astrophysical reaction rate deduced from total cross section deviate from results of BRUSLIB library by factors from two to three in the low temperature region, however, have good consistency nearby T$_{9}$ = 1 GK, which would be an important ingredient for the calculations in nuclear astrophysics.

The authors thank the crew of Pelletron accelerator in the Tokyo Institute of Technology for their help in steady operation of the accelerator. The authors gratefully appreciate financial support from China Scholarship Council. This work has also been supported by the National Natural
Science Foundation of China under Contracts No. 10175091 and No. 11305007 and National Research Foundation of Korea under Contract No. 2018R1A6A06024970.

\appendix


\end{document}